\documentclass[aps,prl,twocolumn]{revtex4}

\usepackage{graphicx}
\usepackage{amsmath,amssymb,latexsym,epsfig}
\usepackage{amssymb}
\usepackage{color}
\usepackage[colorlinks=true, linkcolor=blue, citecolor=blue]{hyperref}

\begin{document} 
\def\be{\begin{equation}}
\def\ee{\end{equation}}

\def\bfi{\begin{figure}}
\def\efi{\end{figure}}
\def\bea{\begin{eqnarray}}
\def\eea{\end{eqnarray}}

\def\figHeight{}

\title{Scaling laws in earthquake occurrence: Disorder, viscosity, and finite size effects in Olami-Feder-Christensen models}

\author{Fran\c cois P. Landes}
\affiliation{The Abdus Salam International Center for Theoretical Physics, Strada Costiera 11, 34014 Trieste, Italy}

\author{E. Lippiello}
\affiliation{Department of Mathematics and Physics, Second University of Naples, Viale Lincoln 5, 81100 Caserta, Italy}

\date{\today}

\begin{abstract}
The relation between  seismic moment and fractured area is crucial to earthquake hazard analysis. Experimental catalogs show  multiple scaling behaviors, with some controversy concerning the exponent value in the large earthquake regime. Here, we show that the original Olami, Feder, and Christensen model does not capture experimental findings. Taking into account heterogeneous friction, the visco elastic nature of faults together with finite-size effects, we are able to reproduce the different scaling regimes of field observations. We  provide an explanation for the origin of the two crossovers between scaling regimes, which are shown to be controlled both by the geometry and the bulk dynamics.
\end{abstract}

\maketitle

The dependence of  the earthquake magnitude $m$ 
on the logarithm of the area $A$ involved in the earthquake fracture process is an outstanding problem of seismic occurrence \cite{Sch82,Rom92,Sch94,WC94,HB02,HB08,Sha09,Sha13,Kon14}. 
This relation  not only provide insights in the mechanisms of earthquake triggering but it is also  necessary in forecasting analyses  to convert predicted slipping areas into expected magnitudes. 
In terms of the seismic moment $M_0\propto 10^{3m/2}$, the ``$m-\log A$'' relation takes the scaling form  $M_0 \propto A^{\eta}$.    

\begin{figure}
\noindent\includegraphics[width=8.7cm]{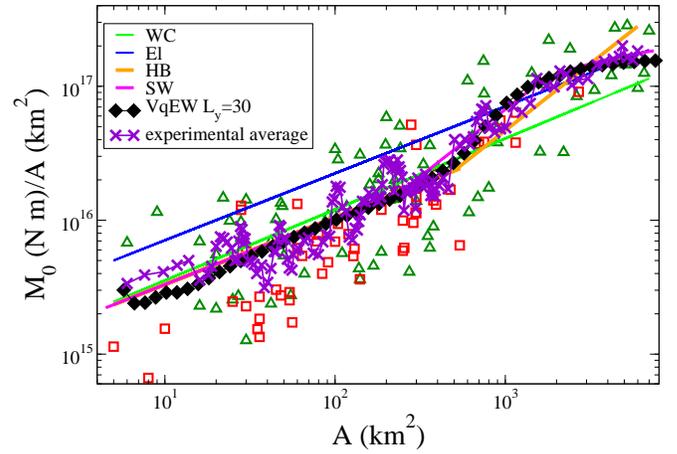}
 \caption{(Color online) The $M_0/A$ versus $A$ scaling relation. 
 Open symbols (each representing an event) are the empirical data set from Table S3 in the electronic supplement of ref.~\cite{Sha13} (green triangles) and  Table 1 of ref.~\cite{Kon14} (red squares).
Continuous lines are different fitting equations (Table \ref{Table}),  
violet cross symbols represent the average value of experimental data and filled black diamonds the result of the VqEW model for $L_y=30$.
We used exponentially increasing bins of width $0.1A$ for this average.
} 
\label{fig1}
\end{figure}

There is general consensus that, for small up to intermediate magnitudes $(m \lesssim 6.5)$, the exponent is $\eta =3/2$, a result well supported by experimental data \cite{WGCEP03}
and some of the conventional models of earthquakes \cite{Esh57,Kno58}. On the other hand, a still open and very debated issue concerns the value of $\eta$ for  ``large'' earthquakes. In this context  earthquakes are defined as ``large'' if $A>A_c \simeq H^2$, where $H$ is the  seismogenic thickness, 
with $H \in [10,25]$ km worldwide. 
More precisely, when the width $W$ of a rectangular slipping area  $A=L \times W$ reaches the thickness of the seismogenic zone $H$, it can only grow along the $L$ direction. 
Under the hypothesis of a constant stress drop (per unit area fractured), conventional models predict $\eta=1$ for $A>A_c$ whereas experimental data indicate larger values $\eta \simeq 2$ \cite{Sch82,Sch94,HB02,HB08} (Fig.~\ref{fig1}). More recent results  interpret the regime  $\eta \simeq 2$ as a crossover before the $\eta=1$ asymptotic regime is recovered \cite{KW07,SW08,Sha09,Sha13}, in agreement with previous observations 
for $A>q A_c$ (with $q\gtrsim 4$) \cite{Rom92}.
The basic problem is that the number of large earthquakes is small. This, combined with uncertainties in the measurement of $A$, makes it very difficult to discriminate different scaling behaviors on the sole basis of experimental data fitting. The breaking of the $M_0$ vs $A$ scaling is expected to produce changes also in the frequency$-$size distribution as soon as the vertical dimension of the earthquake equals $H$ \cite{PSS92}. Nevertheless, the poor statistics of large events does not exclude that observed changes can be an artifact of data analysis \cite{Mai00}.  
At this stage theoretical models thus represent the most efficient way to address this controversial problem. 

In this Rapid Communication, we show how incremental refinements of the single crack model impact and allow to understand the $M_0(A)$ scaling.
Modelling the seismic fault under tectonic drive as a driven interface, and incorporating firstly soft driving, then a random friction force and finally visco-elastic interactions, we are able to identify the origins of the various scaling regimes.
In particular, in the latter case, we are able to reproduce the whole $M_0$ vs $A$ experimental scaling behavior

\begin{table}
\caption{
The values of the exponents $\delta_i$ for 
the most relevant $M_0$ vs $A$ fitting equations. Each fitting equation is named from their proposing reference. We also include the exponents for the three considered numerical models.} 
\begin{tabular}{|l|c|c|c|c|c|c|c|}
\hline
Model  &  WC \cite{WC94} & El \cite{Ell03} & HB \cite{HB08} & SW \cite{SW08}& OFC & qEW & VqEW \\
\hline
$\delta_1$ & 0.03& 0 & 0 & 0 & -0.5 & -0.125 & 0.04 \\
\hline
$\delta_2$ & 0.03& 0 & 0.5 & 0.5 & -0.5 & 0.75 & 0.7 \\
\hline
$\delta_3$ & 0.03& 0 & 0.5 & -0.5 & -0.5 &0.75 & -0.5 \\
\hline
\end{tabular}
\label{Table}
\end{table}

{\it Definitions.}
The seismic moment $M_0$ can be defined as 
\be
M_0 =\mu A \overline{D}
\label{M0}
\ee
where $\mu$ is the shear modulus and $\overline{D}$ is the average displacement within the area $A$.
 Analytical expressions for  $\overline{D}$ in terms of the stress drop $\Delta \sigma$ have been derived for a few specific geometries of slipping areas treated as cracks embedded within homogeneous elastic materials.
These studies \cite{LW95} give $\overline{D}=C \Delta \sigma \Lambda$ where 
$C$ is a factor related to the area's geometry and $\Lambda$ is a characteristic length.
For a rectangular area in particular,  $\Lambda=W$. 
Therefore in the case of space isotropy, $L$ is expected to scale with $W$ ($A \sim W^2$) and $M_0 \propto \Delta \sigma A^{3/2}$. Experimental data at $A<A_c$ are indicating  $\eta=3/2$, consistent with a constant stress drop regime. This leads to  the scaling $\overline{D} \sim L$, supporting a scale invariant behavior where large earthquakes behave as small ones, up to an homogeneous rescaling. The scaling invariance, as well as the spatial isotropy, breaks down when $W \simeq H$. 
In this $A>A_c$ regime, if one keeps the scaling assumption $\Delta \sigma= const$, then  $\overline{D} \sim W \sim L^0$,  $A \sim L$ and $M_0 \sim A$.
Conversely, the scaling behavior $M_0 \sim A^2$ is recovered in the so-called $L$-model, which assumes  $\overline{D} \sim L$, whose mechanical explanation 
conflicts with conventional elastic dislocation theory \cite{Sch94}.

A more complete scaling behavior has been proposed \cite{KW07,SW08,Sha09,Sha13} to interpolate
between $\eta=3/2$ for small earthquakes to $\eta=1$ at large earthquakes, 
according to  the scaling relation 
\be
M_0=A^{\eta_0}F\left(A/A_c\right),
\label{scaling}
\ee
with $\eta_0=3/2$ and 
\be
F(x) \propto \left \{
\begin{array}{l}
 x^{\delta _1}\quad \qquad\mbox{for $x<1$ with $\delta_1=0$ } \\
 x^{\delta _2}\quad \qquad \mbox{for $1<x<q$ with $\delta_2=1/2$ } \\
 q^{\delta _2-\delta_3} x^{\delta_3}\quad \mbox{for $x>q$ with $\delta_3=-1/2$ }
\end{array}
\right .
 .
\label{shaw}
\ee
Under the assumption of a circular area at small $A$ and a rectangular one for $A>A_c$, an estimate $q = 14/3$ was obtained in terms of
the geometric factors $C$ \cite{Sha13}.
In the three regimes the exponent is given by $\eta=3/2+\delta_i$. 
In order to better enlighten the different scaling behaviours we always consider the parametric plots of $M_0/A$ vs $A$ (remember that $M_0/A \sim A^{\eta-1}=A^{\delta+1/2}$). 
In Table \ref{Table} we summarize the values of the exponents $\delta_i$ for the most relevant fitting relations of experimental data. 
In Fig.~\ref{fig1} we plot experimental data of the seismic moment $M_0$ as a function of the area $A$.
The comparison (see Fig.~\ref{fig1}) with the average value of experimental data shows the worst agreement for the single exponent fits (WC,EL) that assume only one fitting parameter and the best agreement for the three exponent fit (SW) that contains two extra fitting parameters. 
    
\begin{figure}
\includegraphics[width=0.5\textwidth]{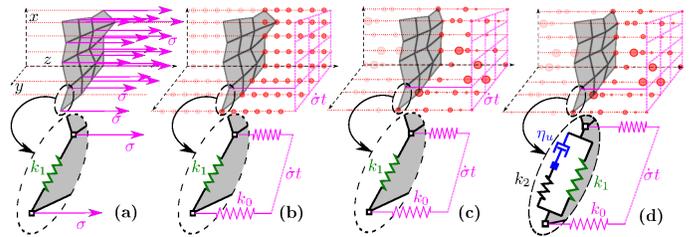}
 \caption{(Color online) 
\textit{Sketches of the four models discussed. }
\textbf{(a)} The classical crack theory.
\textbf{(b)} The OFC model. The pinning forces (red circles) are evenly spaced and have the same strength (proportional to the red circle diamters) in every position $h(x,y)$ of the interface. Interactions are elastic. 
\textbf{(c)} The OFC* or qEW model. The pinning forces are randomly spaced and take random values. 
\textbf{(d)} The VqEW model: in addition, the interactions are visco-elastic.
In panels (b-d) for the sake of clarity the driving spring $k_0$ pulls each site towards the same value $\sigma(t)=\dot{\sigma} t$ and 
the disorder force is pictured only in the $y=0$ and $x=0$ planes; 
\label{fig2}
}
\end{figure}

{\it Dislocation, Crack Theory and the OFC model.}
A sketch of the elastic crack model is presented in Fig.~\ref{fig2}a where the seismic fault (surface of the opened crack) is depicted as blocks interconnected by springs of elastic constant $k_1$ under a constant applied shear stress $\sigma$. 
In this schematic representation $\sigma$ and $\overline{D}$ 
 are pictured along the axis $z$ perpendicular to the fault, 
 despite actually lying within the fault plane.
 This choice allows to represent clearly the continuous renewal of the asperities or pinning forces (red circles in Fig.~\ref{fig2}b-d) 
that happens during sliding, making the random forces $f^{dis}_{x,y}(h(x,y))$ 
independently distributed over  
different fault displacements  $h(x,y)$.

In a more realistic description of seismic occurrence, the effect of tectonic drive is better described by a constant very low shear stress rate $\dot \sigma$, leading to a linearly increasing stress $\sigma(t)=\dot \sigma t$ between rupture events. 
This can be implemented by driving each element of the fault via a spring of elastic constant $k_0$, whose free-end moves at constant velocity (Fig.~\ref{fig2}b). Taking into account a friction force opposing block displacements, one obtains the Burridge-Knopoff (BK) model \cite{BK67}, probably the most simple yet already rich description of a seismic fault. In the OFC model \cite{OFC92}, a cellular automata version of the BK model, the friction term is represented as narrow wells of depth $\sigma_{th}$, so that each block is locked inside a well as long as the applied local stress on the block $\sigma_i$ is less than the threshold $\sigma_{th}$. The model assumes that all wells have the same depth and form a regular lattice (here represented by red dots in Fig.~\ref{fig2}), so that the only random element is the initial distribution of $\sigma_i$'s.

\begin{figure}
\noindent\includegraphics[width=8.7cm]{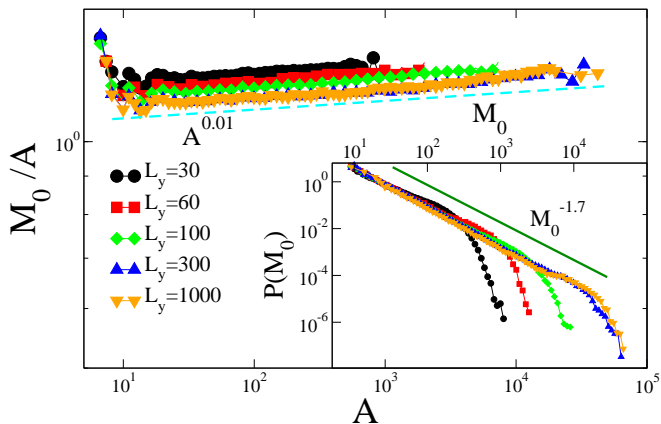}
 \caption{(Color online) Statistical features of avalanches in the OFC model.
 The $M_0$ vs $A$ scaling is in agreement with $\eta  = 1$ in the whole range.
(Inset) The seismic moment distribution $P(M_0)$ is consistent with the power law decay $P(M_0)\sim M_0^{-\tau}$ with $\tau=1.7$.}
\label{fig3}
\end{figure}

The temporal evolution is characterized by stick-slip behavior typical of seismic occurrence, with the periods of quiescence being interrupted only by collective displacements (avalanches). 
Interestingly, for values of  
the elastic coefficients $k_0 \simeq k_1$, 
the seismic moment frequency distribution follows a power law $P(M_0) \sim M_0 ^{-\tau}$, immediately related to the Gutenberg-Richter law for the magnitude distribution, with an exponent $\tau \simeq 1.7$ in very good agreement with experimental data (inset of Fig.~\ref{fig3}) \cite{OFC92}. 
In the case $k_0 \neq 0$, each block involved in an avalanche slips exactly once, leading to  $\overline{D}=\delta h$, where $\delta h$ is the constant inter-well spacing, independent of $A$. The OFC model ($k_1=k_0$) therefore gives $M_0 \propto A$ ($\eta=1$)
 for all values of $A$. This is confirmed by numerical simulations (Fig.~\ref{fig3}) of rectangular faults 
where the length $L_x$ in the direction of the shear stress is kept fixed and sufficiently large to 
   reduce 
finite size effects, whereas different values are considered for the other side $L_y \le L_x=1000$. Free boundary conditions are applied in both directions.

{\it Heterogeneities.}
It has been observed that many phenomena characterized by collective dislocation dynamics, such as plastic deformation as well as vortex (de)pinning in high-$T_c$ superconductors are strongly affected by the presence of various kind of defects \cite{ZZ01,MMZZ04,BFGLV94,OLA15,MVZWG01}.
Indeed, a more realistic description of friction on a fault must also take into account the heterogeneous nature of asperities, which may be controlled by the roughness of the fault, its variable composition, etc.
To that effect, one  can  model friction heterogeneities as narrow potential wells with randomness both in their depth and in their spatial distribution, 
obtaining the so-called OFC* model \cite{Jag10}.  
Concretely, we model the nonlinear term $f^\text{dis}_i(h_i)$, the disorder force, 
 as a series of narrow wells separated by random spacings (Fig.~\ref{fig2}c-d).
The spacings probability density follows the exponential distribution $g(z)= 1/\overline{z} e^{-z/\overline{z}}$ (corresponding to uniformly distributed wells with a density $1/\overline{z}$).
The depth of the wells, which controls the strength of the pinning force in different wells, is taken to be a Gaussian with unit mean and unit variance.
The details of the choice of $g(z)$ and of the strength's distribution is irrelevant at large scales. 
The only thing that matters is that they are both random (finite width of their probability density function) and not fat-tailed (we need to pick short-range correlated distributions for the spacings and the strengths).
We used $\overline{z}=0.1$.

Considering only the main displacements, parallel to the shear direction, OFC* is mapped \cite{AJR12} onto the evolution of an elastic interface driven amongst random impurities, the so-called quenched Edwards-Wilkinson (qEW) universality class \cite{Fisher1998, Kardar1998, Zapperi1998, Giamarchi2006,LeDoussal2013}. 
In this framework, it is natural to establish a relation between the average interface displacement  $\overline{D}$ and the roughness of the interface height profile $h(x,y)$ over a linear length $L$, leading to  $\overline{D} \sim L^\zeta$, where $\zeta$ is the roughness exponent \cite{Zapperi1998, Rosso2003}. 
The exponent can be extracted from the interface structure factor, leading to $\zeta \simeq 1.25$ for one-dimensional interfaces ($d=1$) \cite{Rosso2002b,Ferrero2013a,Ferrero2013} and 
$\zeta \simeq 0.75$ in the $d=2$ case \cite{LeDoussal2002,Rosso2003}. Since 
 $A \sim L^d$, from  Eq.(\ref{M0}) one immediately establishes that
  $\eta=1+\zeta/d$. 
  
\begin{figure}
\noindent\includegraphics[width=8.7cm]{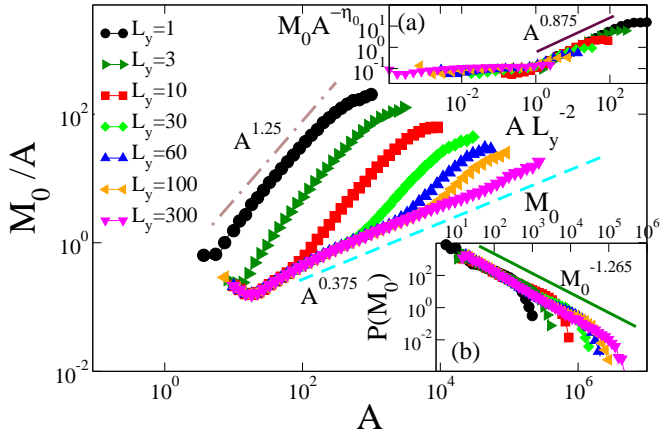}
 \caption{(Color online) Statistical features of avalanches in the qEW (OFC*) model. 
The $M_0$ vs $A$ scaling gives $\eta=2.25$ 
in $d=1$ and $\eta=1.375$ 
 in $d=2$. 
(Inset a) The same data of the main panel with $M_0$ divided by $A^1.375$ plotted versus $A L_y^{-2}$. The maroon continuous line is the $d=1$ power law behavior $A^{2.25-1.375}$.
(Inset b) The seismic moment distribution $P(M_0)$ is consistent with the power law decay $P(M_0)\sim M_0^{-\tau}$ with $\tau=1.265$, in $d=2$. 
}
\label{fig4}
\end{figure}

  This is in agreement with results in Fig.~\ref{fig4} where we plot 
$M_0/A$ vs $A$ for a rectangular fault with 
 $k_0=10^{-4} k_1$, $L_x=1000$ and different values of $L_y$. 
 Periodic boundary conditions are assumed along the shear stress direction $x$.   We find for $L_y=1$ ($d=1$), $\eta=1+\zeta \simeq 2.25$ whereas, in the limit of large $L_y$, we find  $\eta=1+\zeta/2 \simeq 1.375$. 
For intermediate $L_y$ values we observe a crossover from the $d=2$ case when $A<A_c$ ($\eta= 1.375$)
to the $d=1$ case for $A>A_c$, when finite size effects come into play, 
consistently with the scaling behavior Eq.(\ref{scaling}) with $\eta_0=1.375$. 
This is confirmed by the inset (a) of Fig.~\ref{fig4} where, plotting $M_0 A^{-1.375}$ versus $A L_y^{-2}$, we find   a good data collapse  with the scaling function exhibiting the two limiting behaviors $F(x)=const$ for $x \ll1$ and $F(x) \sim x^{2.25-1.375}$ at large x.
The model therefore provides support to the two-exponent fit (HB) with $A_c\sim L_y^2$,  although there is some discrepancy in the value of $\delta_1=-0.125$ (vs $0.0$) and $\delta _2=\delta_3=0.75$ (vs $0.5$, cf.~Table \ref{Table}). We wish to stress that in the finite $L_y$ case the existence of a non-trivial exponent $\zeta>0$ in the scaling relation $\overline{D} \sim L^\zeta$ appears quite naturally in the interface-depinning framework, whereas 
the very existence of a positive roughness $\zeta$
is inconsistent with conventional elastic crack models \cite{Sch82,Sch94}. 

 For the qEW model the roughness exponent also controls the power law decay of $P(M_0)$ and analytical arguments \cite{Narayan1993, LeDoussal2002}  give $\tau=2-2/(\zeta+d)$. This is confirmed by Fig.~\ref{fig4}[inset (b)] where we plot $P(M_0)$ vs $M_0$ for different values of $L_y$ and $k_0/k_1=10^{-4}$. 
In the qEW model the ratio $k_1/k_0$ introduces an upper cut-off $M_c \sim (k_1/k_0)^{(d+\zeta)/2}$ so that $P(M_0) \sim M_0^{-\tau}$ only for $M_0<M_c$, followed by an exponential decay at larger $M_0$.
This is unlike the OFC case where the exponent $\tau$ itself is strongly dependent on the ratio $k_1/k_0$ \cite{OFC92}.
A notable point is that the  $\tau=1.265$ of the qEW model in $d=2$ is significantly smaller than the $\tau \simeq 1.7$ of experimental data.

Therefore, while the introduction of randomly distributed friction forces improves the agreement with experimental data for the $M_0$ vs $A$ scaling, it also makes the agreement for the $P(M_0)$ distribution much worse.

{\it Viscoelastic Interactions.}
Numerical catalogs produced according to the qEW model do not present any burst of activity after large shocks, as opposed to instrumental catalogs.
The introduction of relaxation mechanisms in the inter-avalanche period introduces this ``aftershock'' activity in synthetic catalogs \cite{Jag10,JK10,Jag11,AJR12,Jag13,Jagla2014a,Jagla2014b,Jag14}.
In particular, an elastic coupling between the  fault and a viscoelastic layer (the Asthenosphere) leads to aftershocks with temporal and spatial patterns in very good agreement with experimental data \cite{LGGMD15}. 
In this Rapid Communication we implement visco-elastic relaxation in the bulk, i.e. we consider the simplified form introduced in ref. \cite{Jagla2014a}, which allows for analytic mean field calculations and extensive numerical simulations.

This Viscoelastic qEW (VqEW) model consists in putting in parallel springs $k_1$ with viscoelastic elements built using springs of elastic constant $k_2$ and dashpots as depicted in Fig.~\ref{fig2}d.
The progression of the interface at the point $(x,y) \equiv i$ is denoted $h_i$ and the elongation of the neighbouring dashpots $u_i$.
These fields follow the equations:
\begin{align}
\eta \partial_t h_i
&=  k_0 (V_0 t -h_i) + f_i^\text{dis}(h_i)+k_1 \nabla^2 h _i + k_2 (\nabla^2 h_i - u_i) \notag \\
\eta_u \partial_t u_i
&= k_2 (\nabla^2 h_i - u_i)  ,  \label{2}
\end{align}
where the disorder force is the same as in OFC*. 
Note that there are a priori three time scales in the problem: 
(i) $\tau_D = \overline{z}/V_0$, which accounts for the slow increase of the external drive $w$;
(ii) $\tau = \eta/\max [k_0,k_1,k_2]$, which is the response time of the $h_i$ variables;
(iii) $\tau_u = \eta_u/k_2$, the relaxation time of the secondary field $u_i$. 
In this Rapid Communication we always study the case $\tau \sim 0^+$ (instantaneous avalanches).
Besides, we used $\eta_u=1$ and $V_0=0.0001$ for all simulations, so that for all the values of $k_2 \neq 0$ we used, we have $\tau_u \ll \tau_D$. 
Note that for $k_2=0$, the model reduces exactly to the OFC* model, described by only two time scales $\tau$ and $\tau_D\gg \tau$ (the field $u$ is lost).

\begin{figure}
\noindent\includegraphics[width=8.cm]{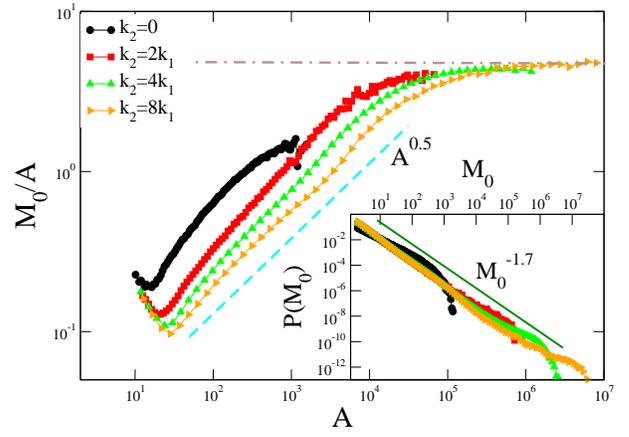}
 \caption{(Color online) 
Statistical features of avalanches in the VqEW model ($k_0=0.06 k_1$).
The $M_0/A$ vs $A$ scaling and the seismic moment distribution $P(M_0)$ (inset) for $L_y=L_x=8000$ and different values of $k_2$. } 
\label{fig5}
\end{figure}

\begin{figure}
\noindent\includegraphics[width=8.7cm\figHeight]{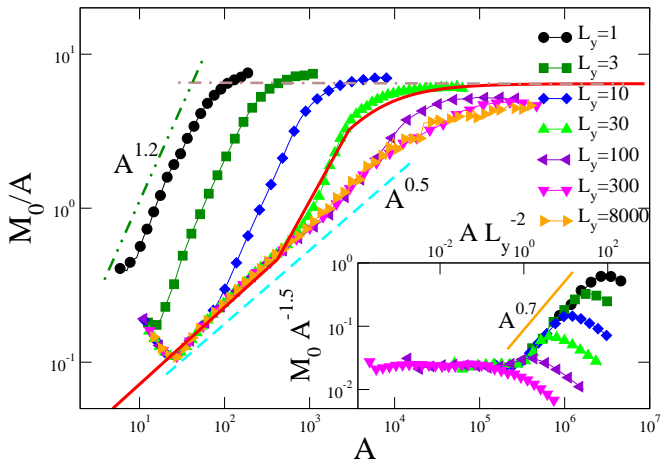}
 \caption{(Color online) 
The $M_0/A$ vs $A$ scaling for $k_2=4 k_1$ and different values of $L_y<L_x=8000$ (and still $k_0=0.06 k_1$). 
The dashed lines represent the fit in the three regimes leading to the three values of $\delta_i$ listed in table \ref{Table}. 
The red continuous line is the scaling form Eq.(\ref{shaw}) obtained as best fit of experimental data \cite{Sha09,Kon14}.
(inset) The same data of the main panel with $M_0$ divided by $A^{1.5}$ plotted versus $A L_y^{-2}$. The continuous green line is the best fit power-law $A^{2.2-1.5}$.}
\label{fig6}
\end{figure}

The relaxation process not only leads to bursts of aftershocks strongly correlated in time and space (similar to the ``migration effect'' \cite{Peng2009}) but also to an avalanche dynamics characterized by new critical exponents, in good agreement with seismic data. 
In particular the exponent $\tau$ changes from $\tau=1.265$ in the qEW model, to $\tau \simeq 1.7$ in the VqEW, a robust result in the limit $k_0 \sim 0^+$ (Fig.~\ref{fig5}, inset) \cite{Jag14,Landes2014Springer}.

In Fig.~\ref{fig5} we explore the $M_0/A$ vs $A$ scaling for different values of $k_2$ in the $d=2$ case, $L_x=L_y=8000$, and  $k_0=0.06 k_1$. Periodic boundary conditions are assumed along the shear stress direction $x$ and we have checked that different choices of  boundary conditions in the $y$-direction do not significantly affect our results.  
We observe that at small $A$, for increasing $k_2$, the  power law exponent changes from $\delta_1=-0.125 \pm 0.05$ 
 when $k_2=0$ (i.e.~back to the qEW model) to a stable value $\delta_1 = 0.04 \pm 0.10$ at larger $k_2$'s. 
Here for $k_2\geq 2k_1$ there exists a characteristic area $A^*$ such that $M_0 \propto A$ ($\eta=1$) for $A>A^*$. 
This crossover area $A^*$ increases with $k_2$, and becomes independent of the system size $L_x=L_y$, as long as it is large enough.
This independence is enlightened by data plotted in Fig.~\ref{fig6}, where different values of $L_y \in [1,8000]$ are considered. Indeed, we observe that 
curves for $L_y \ge 300$ overlap and are numerically indistinguishable. 
Thus, the crossover area $A^*$ can not be due to finite size effects, but emerges from the visco-elastic nature of the model (controlled by $k_2$).

Figure \ref{fig6} shows that for intermediate values of $L_y$ a double crossover pattern is observed, characterized by an intermediate regime
 when $A_c < A < A^*$. 
Unlike $A^*$, the crossover area $A_c$ is mainly controlled by geometric constraints: indeed $A_c$  strongly depends on $L_y$, $A_c \sim L_y^2$. This is confirmed by the inset of fig.~\ref{fig6} where we plot $M_0 A^{-1.5}$ versus $A L_y^{-2}$. When $A<A^*$ data collapse on a scaling function characterized by a flat behavior at small $A$ ($A<A_c$) and a power law behavior $A^{2.2-1.5}$ at larger $A$. 
When $A>A^*$ the 'viscoelastic' regime ($M_0 \sim A$)  sets in and the scaling collapse is violated.  
The intermediate regime $A_c < A < A^*$  can be attributed to a one-dimensional like behavior of the system for finite $L_y$, as confirmed by the the $d=1$ case ($L_y=1$) with $\eta=2.2\pm 0.1$. Conversely, the initial power law $M_0 \sim A^{3/2}$ can be interpreted as a two-dimensional behavior, in agreement with the observation that, for sufficiently large $L_y \ge 300$, $A_c$ becomes larger than $A^*$ and the intermediate regime is not observed. 
The comparison with Fig.~\ref{fig4} indicates that the mechanisms responsible of the first crossover $A< A_c$ are very similar to the elastic case, as confirmed by the fact that  $A_c \sim L_y^2$ in both cases and that $A_c$ is mostly independent of $k_2$. Nevertheless, the presence of the viscoelastic relaxation is still visible in these initial regimes since it affects the value of the exponents $\delta_1$ and $\delta_2$. However the striking effect of the viscoelastic coupling is represented by the asymptotic crossover to $\delta_3= -0.5$ when $A>A^*$, observed for all $L_y$. In correspondence to this crossover,  we also find (inset of Fig.~\ref{fig5}) a small change from $\tau \simeq 1.7$ to a smaller value. The estimate of this asymptotic value of $\tau$, however, can be affected by biases caused by the poor statistics and finite-size effects.

For intermediate $L_y$ values the three-regime pattern we find is 
in good agreement with Eq. (\ref{shaw}) and with experimental data (cf.~Table \ref{Table}).     
This is clearly enlightened in Fig.~\ref{fig1} where we plot results of the VqEW model for $L_y=30$ (after appropriate conversion of arbitrary numerical units). 
For the sake of completeness we also plot in Fig.~\ref{fig6} the scaling function $F(x)$ (Eq.(\ref{shaw})) obtained as best fit of experimental data \cite{SW08} to show that the agreement between numerical results and $F(x)$ extends also beyond the experimental range. 
The VqEW model therefore provides an explanation for the non-trivial scaling 
behavior of $M_0$ vs $A$ observed in instrumental catalogs. 

We interpret the results as follows.
When $A<A_c=L_y^2$, isotropy holds and the $d=2$ exponent $\eta \simeq 1.5$ ($\delta_1 \simeq 0$) is observed. 
For $A_c<A<A^*$ the events reach the boundary 
and avalanches behave as in $d=1$, leading  to
$\eta \simeq 2.2$ ($\delta_2 \simeq 0.7$). 
Finally when  $A>A^*$ the system reaches the full ($d=2$) viscoelastic regime with  $\eta=1.0 \pm 0.025$ ($\delta_3 \simeq -0.5$).
As already observed $A^*$ depends on $k_2$ whereas $A_c$ depends on $L_y$, and for the parameters chosen the experimental value of $q$ (Eq.\ref{shaw})  is recovered for $L_y=30$.

In conclusion, it should be noted that the three regimes and the crossovers between them originate both in the boundary effects and the bulk dynamics, i.e. they cannot be attributed to a single one of these effects.
Our study of the VqEW model accounts for heterogeneous disorder, viscoelastic relaxation mechanisms and finite-size effects, and thus captures all three $M_0(A)$ scaling regimes observed in the field: this accomplishment adds to the previous literature \cite{LGGMD15, Jag14}, which have already shown the relevance of these components for fault models.
Our results promote
 further studies of this class of models, and in particular the investigation of the dependencies of $A^*$ on fault parameters as well as the link between $\eta$ and the local value of the displacement field's roughness $\zeta$.
Inspiration for a better understanding of the large-scale behavior of the VqEW class may come from the OFC model, since both display a robust $\eta=1$ regime, that seems controlled by the strong dissipation rate.

\begin{acknowledgments}
\textit{Acknowledgments ---} 
We thank Alberto Rosso for triggering this collaboration and for his useful suggestions.
\end{acknowledgments}

\end{document}